\documentclass[aps,prd,preprintnumbers,nofootinbib,superscriptaddress,twocolumn]{revtex4}
\usepackage{ae}
\usepackage{bm} 
\usepackage{color}
\usepackage{amsmath}
\usepackage{amssymb}
\usepackage{amsfonts}
\usepackage{graphicx}
\usepackage{slashed}
\usepackage{setspace}
\usepackage{stackrel}
\usepackage{ulem}
\usepackage{upgreek}

\newcommand{\Eqref}[1]{Eq.~\eqref{#1}}

\begin{document}

\title{Vacuum birefringence and diffraction at XFEL: from analytical estimates to optimal parameters}
\author{Elena A. Mosman}\email{mosmanea@tpu.ru}
\affiliation{National Research Tomsk Polytechnic University, Lenin Ave. 30, 634050 Tomsk, Russia}
\author{Felix Karbstein}\email{felix.karbstein@uni-jena.de}
\affiliation{Helmholtz-Institut Jena, Fr\"obelstieg 3, 07743 Jena, Germany}
\affiliation{GSI Helmholtzzentrum f\"ur Schwerionenforschung, Planckstra\ss e 1, 64291 Darmstadt, Germany}
\affiliation{Theoretisch-Physikalisches Institut, Abbe Center of Photonics, \\ Friedrich-Schiller-Universit\"at Jena, Max-Wien-Platz 1, 07743 Jena, Germany}
\date{\today}

\begin{abstract}
We study vacuum birefringence and x-ray photon scattering in the head-on collision of x-ray free electron and high-intensity laser pulses.
Resorting to analytical approximations for the numbers of attainable signal photons, we analyze the behavior of the phenomenon under the variation of various experimental key-parameters and provide new analytical scalings.
Our optimized approximations allow for quantitatively accurate results on the one-percent level.
We in particular demonstrate that an appropriate choice of the x-ray focus and pulse duration can significantly improve the signal for given laser parameters, using the experimental parameters to be available at the Helmholtz International Beamline for Extreme Fields at the European XFEL as example.
Our results are essential for the identification of the optimal choice of parameters in a discovery experiment of vacuum birefringence at the high-intensity frontier.
\end{abstract}

\maketitle

\section{Introduction}

Quantum field theory predicts the quantum vacuum to be characterized by the omnipresence of vacuum fluctuations.
Vacuum fluctuations involving virtual charged particle-antiparticle pairs generically give rise to effective nonlinear couplings between electromagnetic fields, supplementing Maxwell's classical theory of electromagnetism in vacuo with nonlinear interactions.
The high accuracy of classical Maxwell theory for the description of the physics of macroscopic electromagnetic implies that these quantum corrections are very small.
This suggests the possibility of a perturbative expansion of the quantum vacuum nonlinearities in powers of the prescribed electromagnetic fields.

Demanding the effective theory accounting for the vacuum-fluctuation-mediated interactions between macroscopic electromagnetic fields ($\vec{E},\vec{B}$) to be local in the electromagnetic field, to respect Lorentz and gauge invariance, and to exhibit a charge conjugation parity symmetry, the leading nonlinear interaction is a four-field interaction.
More specifically, these assumptions constrain the leading nonlinear correction to classical Maxwell theory to be of the following form ($\hbar=c=1$),
\begin{equation}
 {\cal L}_\text{int}=\frac{m_e^4}{360\pi^2}\biggl[\,a\,\frac{(\vec{B}^2-\vec{E}^2)^2}{4E_{\rm cr}^4} +b\,\frac{(\vec{B}\cdot\vec{E})^2}{E_{\rm cr}^4}\biggr]\,, \label{eq:Lint}
\end{equation}
where $a$ and $b$ denote dimensionless coefficients; cf., e.g., Refs.~\cite{Dittrich:1985yb,Grozin:2020bvd}.
Here, $E_{\rm cr}=m_e^2/e\simeq1.3\times10^{18}\,{\rm V}/{\rm m}$ is the so-called {\it critical} electric field defined in terms of the elementary charge $e$ and the electron mass $m_e$.
The next-to-leading interaction couples six fields and thus is parametrically suppressed with $(E/E_{\rm cr})^2\ll1$.
The explicit values of $a$ and $b$ depend on the details of the underlying quantum field theory, such as, e.g., its particle content.

Specializing to quantum electrodynamics (QED), and accounting for quantum vacuum fluctuations up to two loop order, these dimensionless coefficients become purely numerical and read \cite{Euler:1935zz,Heisenberg:1935qt,Weisskopf:1936kt,Ritus:1975cf}
\begin{equation}
    a=4\Bigl(1+\frac{40}{9}\frac{\alpha}{\pi}\Bigr)\,,
    \quad\quad
    b=7\Bigl(1+\frac{1315}{252}\frac{\alpha}{\pi}\Bigr)\,, \label{eq:a+b}
\end{equation}
where $\alpha=e^2/(4\pi)\simeq1/137$ is the fine structure constant.
Numerous proposals to detect these non-linear effective couplings of macroscopic electromagnetic fields inducing light-by-light scattering phenomena in experiment have been put forward in the literature. Particularly in the past decades, the advent of petawatt-class high-intensity lasers has stimulated research activities in this direction. See, e.g., the recent reviews~\cite{Marklund:2006my,DiPiazza:2011tq,King:2015tba,Karbstein:2019oej}
 and references therein.

One of the most prominent signatures of quantum vacuum nonlinearity in macroscopic electromagnetic fields is vacuum birefringence \cite{Toll:1952}: as a consequence of the effective interaction of electromagnetic fields originally linearly polarized light traversing a strong-field region can pick up a small ellipticity, attributing a birefringence property to the quantum vacuum.
This gives rise to signal photons $N_\perp$ quasi-elastically scattered into an originally empty, perpendicularly polarized mode constituting the signature of quantum vacuum nonlinearity.
Though actively searched for in experiments employing quasi-constant magnetic fields in combination with continuous-wave lasers and high-finesse cavities \cite{Ejlli:2020yhk,Agil:2021fiq,Fan:2017fnd}, so far this effect has never been verified in a controlled laboratory experiment.

\begin{figure*}
\center
\includegraphics[width=0.9\textwidth]{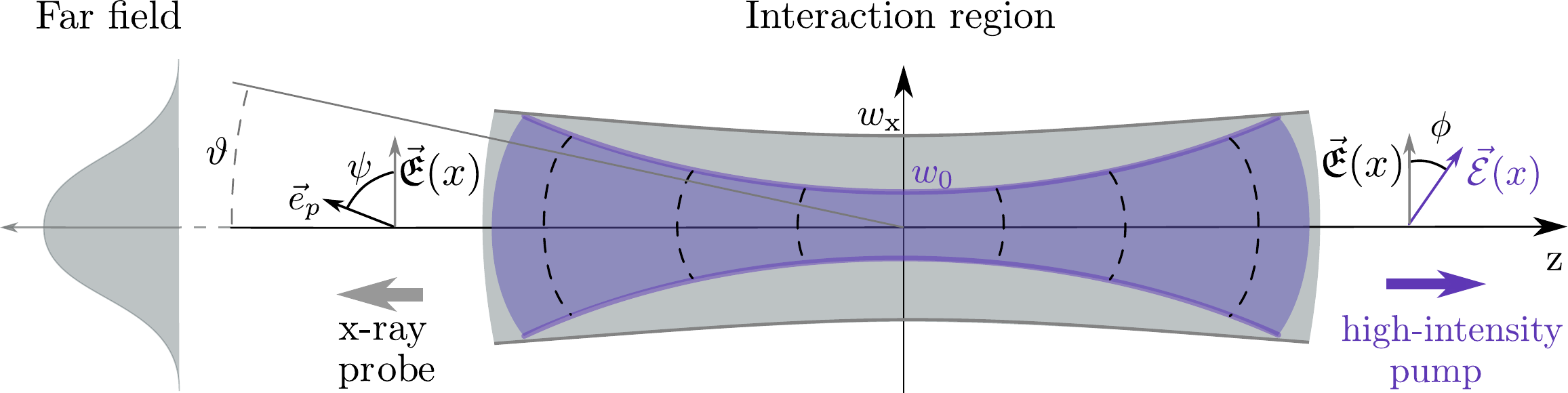}
\caption{Schematic illustration of the considered collision scenario. The x-ray probe~\eqref{eq:Eprobe} propagates in negative $\rm z$ direction and collides head-on with the high-intensity pump~\eqref{eq:E(x)}. Both beams are axially symmetric; $\phi$ is the angle between the polarization vectors of the probe $\vec{\mathfrak E}(x)$ and pump $\vec{\cal E}(x)$ beam. The signals of vacuum birefringence and diffraction are encoded in x-ray signal photons  emitted under an angle of $\vartheta\ll1$ with respect to the forward beam axis of the probe, and to be detected in the far field. The angle between the polarization vectors of the probe $\vec{\mathfrak E}(x)$ and the signal photons $\vec{e}_p$ is $\psi$.}
\label{fig:scenario}
\end{figure*} 

The number of attainable signal photons scales quadratically with the frequency of the probe and the intensity of the pump field, as well as linearly with the number of photons available for probing.
This has triggered theoretical proposals suggesting the use of an x-ray free electron laser (XFEL) as probe and a high-intensity laser as pump; cf., e.g., Refs.~\cite{Aleksandrov:1985,Heinzl:2006xc,DiPiazza:2006pr,King:2013zz,Dinu:2013gaa,Dinu:2014tsa,Schlenvoigt:2016,Karbstein:2015xra,Karbstein:2016lby,Karbstein:2018omb,Karbstein:2020gzg}.
The experimental signature of vacuum birefringence becomes maximum for the head-on collision of the driving laser fields.
In the past years it was in particular emphasized that especially for the scenario involving focused laser beams, even the probe-photon energy preserving signature of vacuum birefringence is generically accompanied by a scattering phenomenon resulting in a different far-field angular decay of the signal photons \cite{Karbstein:2015xra,Karbstein:2016lby,Karbstein:2018omb,Karbstein:2020gzg}. Contrarily to the polarization-flip process, the directional scattering phenomenon itself clearly does not correspond to the key-signature of vacuum birefringence. Hence, such scattering effects were typically neglected  in the theoretical analysis of vacuum birefringence experiments. On the other hand, this quasi-elastic scattering phenomenon is, of course, not limited to the vacuum birefringence signal, but rather amounts to a generic property exhibited by arbitrarily polarized probe photons traversing a spatio-temporally localized strong field region, and thus provides an additional experimental signature of quantum vacuum nonlinearity. See, e.g., Refs.~\cite{Tommasini:2010fb,King:2012aw,Blinne:2018nbd} for the analogous signature in the all-optical frequency domain, Refs.~\cite{dEnterria:2013zqi,Aaboud:2017bwk,Sirunyan:2018fhl,Aad:2019ock} for recent experimental evidences of light-by-light scattering in the ATLAS and CMS experiments at CERN, and Refs.~\cite{Inada:2017lop,Seino:2019wkb} for recent activities towards measuring the x-ray scattering phenomenon in experiment.

So far, no systematic studies of the parameter dependencies of the combined signatures of vacuum birefringence and diffraction for experimentally realistic parameters are available.
Such an analysis is obviously very important for the planning and optimization of the quantum vacuum signatures in upcoming experiments.
Resorting to quantitatively accurate analytical approximations for the differential numbers of signal photons attainable in experiment, in this work we perform a detailed study of various parameter dependencies.
We in particular demonstrate that an appropriate choice of the x-ray focus and pulse duration can significantly improve the signal for given laser parameters, using  the  experimental parameters to be available at the Helmholtz International Beamline for Extreme Fields (HiBEF) at the European XFEL as example. 

Our paper is organized as follows: in Sec.~\ref{sec:analytic_formula} we construct an improved analytic approximation for the differential number of signal photons encoding the signatures of vacuum birefringence and x-ray photon scattering in the head-on collision of an XFEL probe with a high-intensity laser pump.
In Sec.~\ref{sec:parameters} we detail our assumptions for the parameters characterizing the experimental parameters available at HiBEF at the European XFEL.
Subsequently, in Sec.~\ref{sec:results} we introduce three different experimental observables for vacuum birefringence and x-ray photon scattering, and in detail discuss their dependencies on the parameters of the driving laser fields. Our main focus is on the question of how to optimize the signal for given experimental parameters.
Finally, we end with conclusions and a brief outlook in Sec.~\ref{sec:Concls}.

\section{Signal photon numbers}\label{sec:analytic_formula}

Throughout this work, we consider the head-on collision of a loosely focused x-ray probe with a tightly focused optical high-intensity laser pump at zero impact parameter; see Fig.~\ref{fig:scenario} for a graphical illustration.
Both beams are expected to be well-described as linearly polarized paraxial Gaussian beams, endowed with a temporal Gaussian pulse envelope.
The field profile of the pump laser field is \cite{Siegman}
\begin{align}
 {\cal E}(x)=&{\cal E}_0\,{\rm e}^{-\frac{({\rm z}-t)^2}{(\tau/2)^2}}\frac{w_0}{w({\rm z})} \,{\rm e}^{-\frac{r^2}{w^2({\rm z})}} \label{eq:E(x)}\\ \notag
&\times\cos\bigl(\Omega({\rm z}-t)+\tfrac{\Omega r^2}{2R({\rm z})}-\arctan(\tfrac{\rm z}{{\rm z}_R})\bigr)\,, 
\end{align}
with peak field amplitude ${\cal E}_0$, oscillation frequency $\Omega$, and pulse duration $\tau$; $r$ is the radial coordinate.
The beam radius $w({\rm z})=w_0\sqrt{1+({\rm z}/{\rm z}_R)^2}$ describes the widening of the beam as a function of the longitudinal coordinate $\rm z$; $w_0$ is the beam waist and ${\rm z}_R=\pi w_0^2/\lambda$ is the Rayleigh length. $R({\rm z})={\rm z}\bigl[1+({\rm z}_R/{\rm z})^2\bigr]$ denotes the radius of curvature of the wavefronts and $\arctan({\rm z}/{\rm z}_R)$ is the Gouy phase.
Adopting the field profile~\eqref{eq:E(x)}, the deviations from the corresponding exact solution of Maxwell's equations in vacuo are parametrically suppressed by positive powers of $1/(\Omega\tau)\ll1$ and $\epsilon:=w_0/{\rm z}_R<1$ \cite{Karbstein:2017jgh}. However, resorting to the results of Refs.~\cite{Davis:1979,Barton:1989,Salamin:2007} it can be explicitly shown that in the considered pump-probe experiment corrections beyond the leading order of the paraxial approximation for the pump field effectively affect the results for the signal photon numbers only at ${\cal O}(\epsilon^2)$ \cite{KarbsteinSundqvist:2021}.

As it is focused more loosely, i.e., fulfills $2/(\omega w_{\rm x})\ll1$, for the x-ray probe of photon energy $\omega$ and beam waist $w_{\rm x}$ it is justified to use the infinite Rayleigh length approximation: the longitudinal extent of the interaction volume of the colliding beams is controlled by the much shorter Rayleigh range of the optical laser beam. On this scale, the beam radius of the x-ray probe increases from $w_{\rm x}$ to $w_{\rm x}({\rm z}_R)=w_{\rm x}\sqrt{1+(\Omega/\omega )^2( w_0/w_{\rm x})^4}$, implying that the infinite Rayleigh length approximation yields trustworthy results in the parameter regime where $(w_0/w_{\rm x})^2\,\Omega/\omega\ll1$.
Correspondingly, the probe field profile can be expressed as
\begin{equation}
 {\mathfrak E}(x)={\mathfrak E}_0\,{\rm e}^{-\frac{({\rm z}+t)^2}{(T/2)^2}}{\rm e}^{-\frac{r^2}{w_{\rm x}^2}}\,
  \cos\bigl(\omega({\rm z}+t)\bigr) , \label{eq:Eprobe}
\end{equation}   
with pulse duration $T$, and peak field amplitude ${\mathfrak E}_0$.
For this field, the restriction on the pulse durations to be reliably considered reads $1/(\omega T)\ll1$. We have explicitly ensured, that for all the parameter sets discussed in this work the above conditions are met.
The peak field amplitudes can be expressed in terms of the respective laser pulse energy, beam waist and pulse duration. 
This yields \cite{Karbstein:2017jgh} 
\begin{equation}
{\cal E}_0^2=8\sqrt{\frac{2}{\pi}}\frac{W}{\pi w_0^2\tau}\,, \qquad {\mathfrak E}_0^2=8\sqrt{\frac{2}{\pi}}\frac{N \omega}{\pi w_{\rm x}^2 T}\,,
\end{equation}
where $W$ is the laser pulse energy of the high-intensity laser and $N$ is the number of photons constituting the x-ray beam.
The far-field angular decay of the latter as function of the polar angle $\vartheta\ll1$ measured from the forward beam axis is described by \cite{Karbstein:2019oej,Siegman}
\begin{equation}
    \frac{{\rm d}N}{\vartheta\,{\rm d}\vartheta}\simeq N(\omega w_{\rm x})^2\,{\rm e}^{-\frac{1}{2}(\omega\vartheta w_{\rm x})^2} \,. \label{eq:decay}
\end{equation}

The superposition of the probe and pump laser pulses gives rise to signal photons of polarization $p$ and wavevector $\vec{k}={\rm k}(\cos\varphi\sin\vartheta,\sin\varphi\sin\vartheta,\cos\vartheta)$ encoding the signature of quantum vacuum nonlinearity in experiment \cite{Karbstein:2014fva,Karbstein:2019oej}.
The experimental signatures of vacuum birefringence and diffraction are x-ray signal photons of energy ${\rm k}\simeq\omega$, which for kinematic reasons are predominantly emitted into the forward cone of the probe laser beam, such that generically $\vartheta\ll1$.
In this limit, the polarizations of the signal photons can be spanned by the vector
$\vec{e}_p\simeq(\cos\psi,\sin\psi,0)$ and thus be parameterized by the single angle parameter $\psi$. Choosing the polarization vector of the probe without loss of generality as $\vec{\mathfrak E}(x)\sim(1,0,0)$, an angle of $\psi=\pi/2$ corresponds to polarization flipped signal photons, such that $p\to\,\perp$. On the other hand $\psi=0$ would correspond to signal photons polarized parallel to the probe, i.e., $p\to\,\parallel$.
For the head-on collision of the driving laser pulses at zero impact parameter as considered here, the signal moreover exhibits a rotational symmetry about the beam axis, i.e., does not explicitly depend on the polar angle $\varphi$.
In turn, all the nontrivial information about the signal is encoded in the differential number ${\rm d}N_p/{\rm d}\!\cos\vartheta\simeq{\rm d}N_p/(\vartheta\,{\rm d}\vartheta)$. See Fig.~\ref{fig:scenario} for an illustration.

Accounting for generic polarizations of the signal \cite{Karbstein:2019bhp} and specializing to the rationally symmetric case for the x-ray probe at zero impact parameter, the analytical approximation introduced in 
Ref.~\cite{Karbstein:2018omb} can be expressed as
\begin{align}
\frac{{\rm d}N_p}{\vartheta\,{\rm d}\vartheta} 
 &\simeq
\bigl[c_+\cos\psi+c_-\cos(\psi-2\phi)\bigr]^2\frac{4\alpha^4}{225(3\pi)^\frac{3}{2}} \nonumber\\
&\quad\times N\Bigl(\frac{W}{m_e}\frac{\lambdabar_{\rm C}}{w}\Bigr)^2\Bigl(\frac{w}{w_0}\frac{\omega}{m_e}\Bigr)^4
\frac{(\frac{w_{\rm x}}{w})^2}{[1+2(\frac{w_{\rm x}}{w})^2]^2}\nonumber\\
&\quad\times{\rm e}^{-\frac{1}{2}\frac{(\omega\vartheta w_{\rm x})^2}{1+2(\frac{w_{\rm x}}{w})^2}} F\Bigl(\tfrac{4{\rm z}_R}{\sqrt{T^2+\frac{1}{2}\tau^2}},\tfrac{T}{\tau}\Bigr)\,, \label{eq:d2NF}
\end{align}
with numerical coefficients $c_\pm=a\pm b$, and
\begin{align}
F(\chi,\rho):=&\sqrt{\frac{1+2\rho^2}{3}}\,\chi^2\, {\rm e}^{2\chi^2} \, \int_{-\infty}^\infty {\rm d}\kappa\,{\rm e}^{-\kappa^2}\,\notag\\&\times\biggl|\sum_{\ell=\pm1} {\rm e}^{2\ell\rho\kappa\chi} 
 {\rm erfc}\bigl(\ell\rho\kappa
 +\chi\bigr)\biggr|^2 \,. \label{eq:F}
\end{align}
See Ref.~\cite{Karbstein:2018omb} for analytic expressions of the function $F(\chi,\rho)$ in various limiting cases.
Here, $w\gtrsim w_0$ denotes the effective waist of the pump beam, which characterizes the transverse extent of the interaction volume.

This approximation was derived under the following critical assumptions. First, the factor of $w({\rm z})$ in the exponential of \Eqref{eq:E(x)} is replaced by the effective waist $w$. Second, all manifestly inelastic contributions to the signal exhibiting an explicit dependence on the pump laser frequency $\Omega$ are dropped. Third, an expansion in $\vartheta\ll1$ was performed: the leading neglected correction to the differential signal photon number scales as $T\omega\vartheta^2\ll1$.
For the details, see Refs.~\cite{Karbstein:2016lby,Karbstein:2018omb}.

Reference~\cite{Karbstein:2018omb} fixed the effective waist by averaging the beam radius $w({\rm z})$ over one Rayleigh length, yielding $w\simeq1.15w_0$.
Adopting this choice, a reasonable agreement with the results of a full numerical integration \cite{Karbstein:2016lby} for the same parameters was demonstrated.
For the different cases considered there, the relative deviations in the integrated signal photon numbers were below $15\%$; see Table~I of Ref.~\cite{Karbstein:2018omb}.

In the present work, we use a more advanced strategy to fix the effective waist $w$, namely we choose it such that that for $\vartheta=0$ the approximate result for the differential number of signal photons~\eqref{eq:d2NF} matches the result of an exact evaluation of the same quantity; the latter is stated in Eqs.~(19)-(21) of Ref.~\cite{Karbstein:2016lby}.  
Remarkably, for $\vartheta=0$ this exact result can be integrated analytically over the longitudinal coordinate for axially symmetric beams colliding at zero impact parameter; it can be expressed in terms of function defined in \Eqref{eq:F}.
This identification results in the following identity,
\begin{align}
  \biggl(\frac{1+2(\frac{w_{\rm x}}{w})^2}{1+2(\frac{w_{\rm x}}{w_0})^2}\biggr)^2=\,\frac{F\Bigl(\tfrac{4{\rm z}_R}{\sqrt{T^2+\frac{1}{2}\tau^2}},\tfrac{T}{\tau}\Bigr)}{F\Bigl(\tfrac{4{\rm z}_R\sqrt{1+2(\frac{w_{\rm x}}{w_0})^2}}{\sqrt{T^2+\frac{1}{2}\tau^2}},\tfrac{T}{\tau}\Bigr)} \,, \label{eq:cond}
\end{align}
which can be readily solved for $w$.

Equation~\eqref{eq:cond} implies that the effective waist is fully determined by geometric properties: it is a function of the pulse durations and waist sizes of both the pump and the probe beams as well as the Rayleigh length of the pump.
We emphasize the difference to the naive, solely $w_0$-dependent choice of $w\simeq1.15w_0$ in all parameter regimes.

In a wide range of parameters the effective waist defined by \Eqref{eq:cond} indeed turns out to vary in the range $1\lesssim w/w_0\lesssim 1.15$.
However, specifically for small ${\rm z}_R$ and large $w_{\rm x}$, also significant deviations from these values are possible.
See Tab.~\ref{tab:vgl} for a comparison of the accuracy of our new identification and the choice of $w\simeq1.15w_0$ adopted previously \cite{Karbstein:2018omb}.
\begin{table}
 \caption{Relative deviation $|1-N_\perp/N^\text{full}_\perp|$ of the numbers of polarization-flipped signal photons $N_\perp$ obtained from the approximation~\eqref{eq:d2NF} and the corresponding exact results $N^\text{full}_\perp$ obtained in Ref.~\cite{Karbstein:2016lby}.
 Here, we compare the accuracy of the naive choice of $w\simeq1.15w_0$ for the effective waist and our new advanced identification via \Eqref{eq:cond} for different choices of the probe waist $w_{\rm x}$; $\lambda=800\,{\rm nm}$, $\tau=T=30\,{\rm fs}$, $w_0=1\upmu{\rm m}$, $\omega=12914\,{\rm eV}$.}
\label{tab:vgl}
\begin{ruledtabular}
\begin{tabular}{ccc}
 $w_{\rm x}/w_0$ & $w/w_0=1.15$ & \Eqref{eq:cond} \\
\colrule
$1/10$ & $2.6\%$ & $0.9\%$   \\
  $1/3$ & $0.2\%$ & $0.9\%$ \\
  $1$ & $8.0\%$ & $0.5\%$ \\
  $3$ & $13.8\%$ & $0.1\%$  \\
\end{tabular}
\end{ruledtabular}
\end{table}

Table~\ref{tab:vgl} clearly showcases the superiority of our new identification: for all considered cases the relative deviation of the values determined with our analytical approximation and those obtained from a full numerical evaluation is below $1\%$.

Some clarifications are in order here. The replacement $w({\rm z})\to w$ in the exponential of the pump field profile~\eqref{eq:E(x)} should certainly be justified if either $\tau\ll{\rm z}_R$ or $T\ll{\rm z}_R$: the signal is predominantly originating from the region where the electromagnetic field is maximal, which in this limit is restricted to the region where $w({\rm z})\simeq w_0$. In addition, this replacement should be perfectly justified for the case of $w_{\rm x}\ll w_0$: in this limit the dependence of the transverse pump profile drops out and the probe photons only see the on-axis profile of the pump which is consistently taken into account. Finally, the fact that the replacement works well also for $w_{\rm x}\gtrsim w_0\approx w$ is nicely exemplified by the values for $w_{\rm x}/w_0=\{1,3\}$ given in Tab.~\ref{tab:vgl}. Noteworthy, the increase in the relative deviation observed for decreasing values of $w_{\rm x}/w_0$ in Tab.~\ref{tab:vgl} is not due to the introduction of the effective waist. The reason for this increase is rather the grow of the sub-leading corrections $\sim T\omega\vartheta^2$ for $w_{\rm x}/w_0\to 0$: in this limit the radial divergence of the signal is approximately given by $2/(\omega w_{\rm x})$ \cite{Karbstein:2018omb}, such that these corrections scale as $\sim T/(\omega w_{\rm x}^2)$. For $w_{\rm x}/w_0=1/10$ and the parameters of Tab.~\ref{tab:vgl} this factor becomes as large as $\simeq 0.01$.

Hence, in combination with \Eqref{eq:cond}, \Eqref{eq:d2NF} allows for a very accurate description of the differential number of signal photons, and -- essentially without loss of precision -- can be used interchangeably with the analogous result of a full numerical evaluation. In fact, the small encountered deviations one the $1\%$ level are comparable to the error margin of the leading paraxial approximation. However, due to both its analytical nature and relative simplicity, our expression allows for the transparent tracing of parameter dependencies, and can even be straightforwardly differentiated.
This opens up new opportunities, such as identifying the optimal parameters maximizing the effect, while analytically accounting for additional experimental parameter dependencies.

We also note that Ref.~\cite{Karbstein:2018omb} considered the pump laser field to be focused down to $w_0=1\upmu{\rm m}$, such that in this case the parameter controlling the applicability of the paraxial approximation for the pump reads $\epsilon^2\simeq0.06$. Conversely, throughout the present work we assume $w_0=1.7\upmu{\rm m}$, which implies $\epsilon^2\simeq0.02$. In line with that, particularly for the latter choice we expect the results of the present article to be quantitatively accurate on the few percent level.

For convenience, in the remainder of this work we employ the following shorthand notations:
\begin{align}
\beta:=\frac{w_{\rm x}}{w_0}\,\quad\text{and}\quad F_{\beta}:= F\Bigl(\tfrac{4{\rm z}_R\sqrt{1+2\beta^2}}{\sqrt{T^2+\frac{1}{2}\tau^2}},\tfrac{T}{\tau}\Bigr)\,. \label{eq:shct}
\end{align}
Besides, we limit our discussion to the number of polarization-flipped signal photons $N_\perp$ and the total number of signal photons attainable in a polarization insensitive measurement $N_{\rm tot}$.
As noted above, the former is obtained by setting $\psi=\pi/2$, and the latter follows upon summing over two perpendicular signal polarizations, i.e., adding the results for $\psi$ and $\psi\to\psi+\pi/2$.

Explicitly accounting for \Eqref{eq:cond}, using the notations~\eqref{eq:d2NF} and specializing to $N_\perp$ and $N_{\rm tot}$, \Eqref{eq:d2NF} can be recast as
\begin{align}
\left\{{\rm d}N_{\rm tot} \atop {\rm d}N_\perp\right\}
 &\simeq \vartheta\,{\rm d}\vartheta
\left\{2(a^2+b^2)+2(a^2-b^2)\cos(2\phi) \atop (a-b)^2\sin^2(2\phi)\right\} \nonumber\\
&\quad\times\frac{4\alpha^4}{225(3\pi)^\frac{3}{2}}\,
N\Bigl(\frac{W}{m_e}\frac{\lambdabar_{\rm C}}{w_0}\Bigr)^2\Bigl(\frac{\omega}{m_e}\Bigr)^4 \nonumber\\
&\quad\times\frac{\beta^2}{(1+2\beta^2)^2}\,{\rm e}^{-\frac{1}{2}\frac{(\omega\vartheta w_0 \beta)^2}{1+2\beta^2}\sqrt{\frac{F_\beta}{F_0}}} \,F_\beta \,. \label{eq:d2N}
\end{align}
For completeness, we note that these results can be readily generalized to collisions at a finite impact parameter: to this end the effective waist in Eq.~(4) of Ref.~\cite{Karbstein:2018omb} is to be identified with $w$ in \Eqref{eq:cond}.
This in particular implies that the differential signal photon numbers scale as
\begin{equation}
 \sim{\rm e}^{-4(\frac{r_0}{w})^2\bigl[1+2(\frac{w_{\rm x}}{w})^2\bigr]^{-1}}={\rm e}^{-4(\frac{r_0}{w_{\rm x}})^2}\Bigl(1+{\cal O}\bigl((\tfrac{w}{w_{\rm x}})^2\bigr)\Bigr) \label{eq:impact}
\end{equation}
with a finite transverse impact parameter $r_0$.

Finally, we emphasize that the measurement of the signal photon numbers in two different polarization states, such as $N_{\rm tot}$ and $N_\perp$ for given $\phi$, or $N_{\rm tot}$ for two different relative polarizations of the driving beams $\phi$, allows for the individual extraction of the coefficients $a$ and $b$ in \Eqref{eq:Lint}. Standard birefringence experiments only provide access to the difference $a-b$; cf. the second line of \Eqref{eq:d2N}.

Presuming that QED yields the dominant contribution, subsequently we adopt the QED predictions for $a$ and $b$ in \Eqref{eq:a+b} and study how to achieve an optimal signal for the possible parameter available at HiBEF at the European XFEL.

\section{Experimental parameters available at the European XFEL}\label{sec:parameters}

The numbers of probe photons $N$ per pulse available at the European XFEL depend on the desired photon energy $\omega$, as well as the pulse duration $T$ employed for the experiment \cite{Schneidmiller:95609}.
This dependency was not accounted for in previous studies.
To achieve the largest possible values for $N$, we choose the highest possible FEL electron energy of $17.5\,{\rm GeV}$.
The relevant parameters for HiBEF, which is located at the High Energy Density instrument, are
detailed in Tables C.1-C.6 of Ref.~\cite{Schneidmiller:95609}.

Aiming at the study of polarization-flipped signal photons we moreover choose a probe photon energy of $\omega=12914\, {\rm eV}$. For this energy the possibility of high-definition polarimetry was successfully demonstrated in experiment using Bragg reflections at silicon crystals \cite{Marx:2013xwa}, and the present polarization purity record of ${\cal P}\simeq1.4 \times 10^{-11}$ was achieved with four reflections.
See Tab.~\ref{tab:times} for the XFEL photon numbers for various possible pulse durations at precisely this photon energy.

Another important issue is that the channel-cut polarizer employed to achieve the high-definition polarization state of the x-ray beam needed for a vacuum birefringence experiment generically increases the pulse duration of the x-ray beam. 
The corresponding increase $T^\text{FWHM}\to T_{\cal P}^\text{FWHM}$ can be estimated along the lines of Refs.~\cite{Lindberg:2012tn,Shvydko:2012rzc}; cf. Ref.~\cite{Karbstein:2021ldz} for more details.
Assuming Gaussian temporal pulse profiles before\footnote{The temporal structure of realistic XFEL pulses is typically not smooth at all, but rather very complicated and characterized by a large number of spikes; see, e.g., Fig.~7 in \cite{Schneidmiller:95609}.
Noteworthy, after a few reflections the resulting pulse profile is substantially smoothed and closely resembles a Gaussian pulse profile.}
and after the four reflections in the channel-cut we obtain the values of the pulse durations $T_{\cal P}^\text{FWHM}$ listed in Table \ref{tab:times}.
We emphasize that this implies that for studies of the polarization-flipped signal the probe pulse duration has to be identified with $T_{\cal P}$; the original XFEL pulse durations $T$ are only available for polarization insensitive measurements.
\begin{table}[b]
\caption{Numbers of probe photons $N$ per pulse available at the European XFEL as a function of the {\it full width at half maximum} pulse duration $T^\text{FWHM}$ for a photon energy of $\omega=12914\,{\rm eV}$.
A polarizer generically modifies the incident pulse duration as $T^\text{FWHM}\to T_{\cal P}^\text{FWHM}$.
In the third column we give the corresponding pulse durations $T_{\cal P}^\text{FWHM}$ after traversing a four-reflection silicon channel-cut polarizer.}
\label{tab:times}
\begin{ruledtabular}
\begin{tabular}{lccr}
\# &$T^\text{FWHM} [{\rm fs}]$ & $T_{\cal P}^\text{FWHM} [{\rm fs}]$ & $N\qquad $ \\
\colrule
(1)& $1.67$ & $100.1$ & $2.98 \times 10^{10}$ \\
(2)& $8.96$ & $100.3$ & $1.78 \times 10^{11}$ \\
(3)& $23.2$ & $101.4$ & $3.22 \times 10^{11}$ \\
(4)& $42.8$ & $104.5$ & $5.03 \times 10^{11}$ \\
(5) & $107$ & $129   $ & $8.26 \times 10^{11}$ \\
\end{tabular}
\end{ruledtabular}
\end{table}

For the high-intensity laser pump we adopt the parameters of the $300\,{\rm TW}$ {\it Relativistic Laser at XFEL} (ReLaX) system installed at HiBEF, delivering pulses of energy $W=10\,{\rm J}$ and duration $\tau^\text{FWHM}=25\,{\rm fs}$ at a wavelength of $\lambda=800\,{\rm nm}$, focused to a waist size of $w_0=1.7{\rm \upmu m}$ (HWHM value $1\upmu{\rm m}$).
Note that FWHM pulse durations are related to $1/{\rm e}^2$ pulse durations as $\tau\simeq 1.7\,\tau^{\rm FWHM}$.

For completeness and to allow for a simple comparison, we also note that the projected parameters of the high-intensity laser at SACLA are somewhat more challenging, but otherwise quite similar to those at HiBEF, namely $W=12.5\,{\rm J}$ in $\tau^\text{FWHM}=25\,{\rm fs}$ at $\lambda=800\,{\rm nm}$ focused to $w_0=1{\rm \upmu m}$. Another difference is the probe frequency of $\omega=9.8\,{\rm keV}$ envisioned for the SACLA experiment, aiming exclusively at an polarization insensitive measurement of vacuum diffraction with $N\simeq3\times10^{11}$ XFEL photons per pulse \cite{Seino:2019wkb}.

\section{Results}\label{sec:results}

In the present work we consider two different choices for the relative polarization $\phi$ of the pump and probe laser beams \cite{Karbstein:2019bhp}. 
The total number of signal photons attainable in a polarization insensitive measurement $N_{\rm tot}$ is maximized for $\phi=\pi/2$ and the number of polarization-flipped signal photons for $\phi=\pi/4$.
We stick to these two choices in the following, and provide results for $N_{\rm tot}|_{\phi=\pi/2}\to N_{\rm tot}$ and $N_\perp|_{\phi=\pi/4}\to N_\perp$ only.

Subsequently, we analyze the parameter dependence of the signal accessible by three different observables: the integrated number of polarization-flipped signal photons $N_\perp$ in Sec.~\ref{subsec:a}, the discernible number of polarization-flipped signal photons $N_{\perp>}$ in Sec.~\ref{subsec:b}, and the total number of discernible signal photons attainable in a polarization insensitive measurement $N_{{\rm tot}>}$ in Sec.~\ref{subsec:c}. Noteworthy, the parameter dependencies of the considered observables are quite distinct.

\subsection{Integrated number of signal photons}\label{subsec:a}

\begin{figure}
\center
\includegraphics[width=0.45\textwidth]{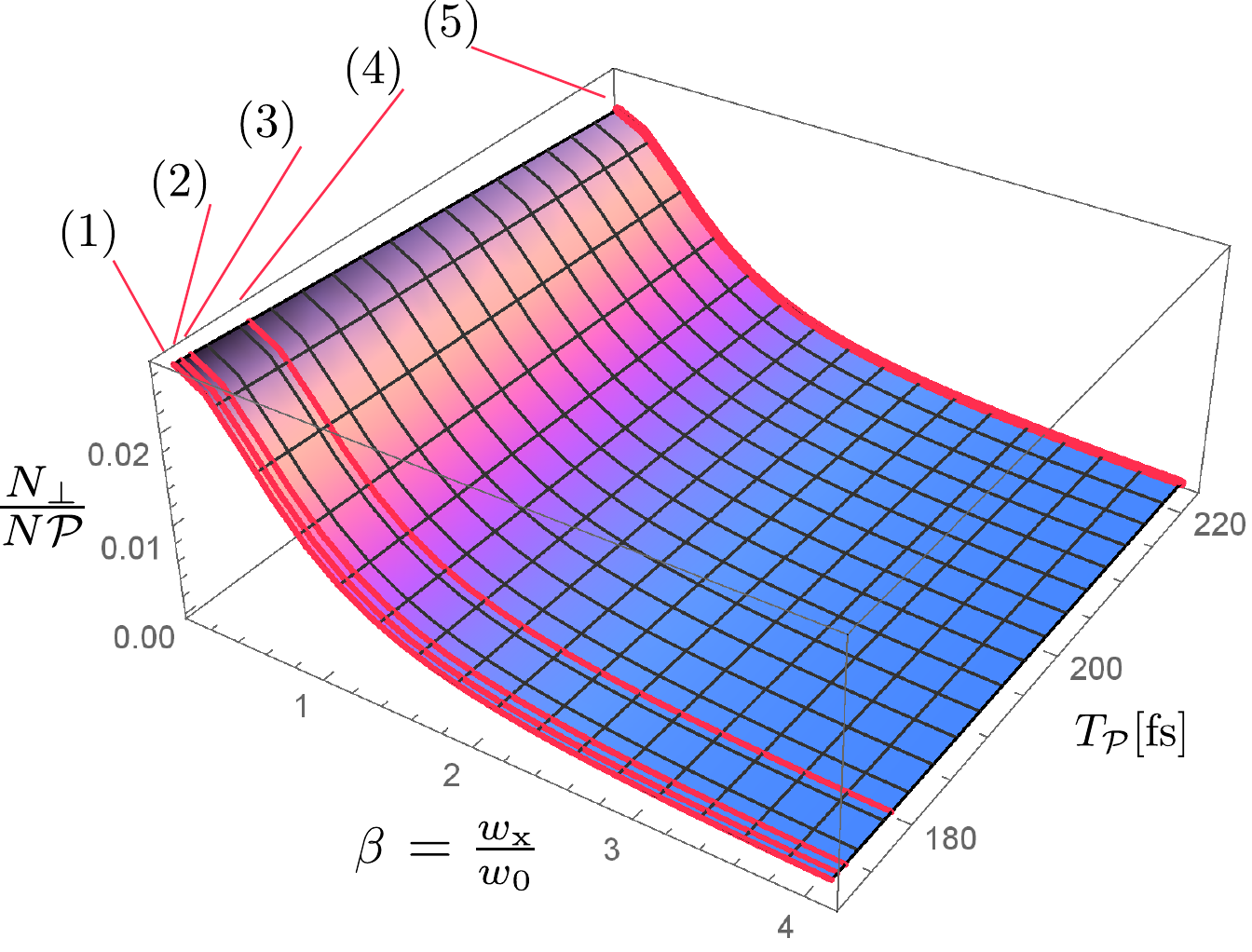}
\caption{Dependence of $N_\perp/({\cal P}N)$ on the probe pulse duration $T$ and probe waist $\beta=w_{\rm x}/w_0$ for $\omega=12914\,{\rm eV}$ and ${\cal P}=1.4 \times 10^{-11}$. The pump laser parameters are those available at HiBEF: $W=10\,{\rm J}$, $\tau=42\,{\rm fs}$, $\lambda=800\,{\rm nm}$, $w_0=1.7\upmu{\rm m}$. The red contours (1)-(5) mark the results obtained for the probe pulse durations listed in Tab.~\ref{tab:times}; for intermediate times we use a smooth monotonic interpolation $N(T)$ of the these values.}
\label{fig:full}
\end{figure}

The ratio of the integrated number of signal photons and the number of x-ray photons available for probing can be expressed as
\begin{align}
\frac{N_{\perp}}{N}
 &\simeq \frac{4\alpha^4}{25(3\pi)^\frac{3}{2}}\,\Bigl(\frac{W}{m_e}\frac{\omega}{m_e}\Bigr)^2\Bigl(\frac{\lambdabar_{\rm C}}{w_0}\Bigr)^4\,
\frac{1}{1+2\beta^2} \,\sqrt{F_\beta F_0}\,. \label{eq:NperpbyN}
\end{align}
This result follows upon integration of the second line of \Eqref{eq:d2N} over the polar angle $\vartheta$.

In accordance with elementary plane-wave considerations predicting a scaling of the signal photon number $\sim\omega^2{\cal E}_0^4\sim\omega^2W^2/(w_0^4\tau^2)$ \cite{BialynickaBirula:1970vy,Baier,Brezin:1971nd}, \Eqref{eq:NperpbyN} scales quadratically with both the probe photon energy $\omega$ and the pulse energy of the pump $W$.
However, in general the scaling  of \Eqref{eq:NperpbyN} with the pump waist size $w_0$ shows a slight deviation from the above plane-wave prediction, and $w_0^4N_\perp/N$ still increases with $w_0$.
The factor $\sqrt{F_\beta F_0}$ slowly declines with increasing both $T$ and $\tau$, such that the minimal possible values for $T$ and $\tau$ maximize the birefringence signal.
Note, that all the parameters on the right-hand side of \eqref{eq:NperpbyN} are independent of each other.

In Fig.~\ref{fig:full} we highlight the behavior of $N_\perp/(N{\cal P})$ under variations of the pulse duration $T$ and waist $w_{\rm x}$ of the x-ray pulse.

One can see that the maximum is reached for $\beta \ll1$. In this limit \Eqref{eq:NperpbyN} reproduces Eq.~(8) of \cite{Karbstein:2018omb},
\begin{align}
\frac{N_{\perp}}{N {\cal P}}
 &\simeq \frac{4\alpha^4}{25(3\pi)^\frac{3}{2}}\,\Bigl(\frac{W}{m_e}\frac{\omega}{m_e}\Bigr)^2\Bigl(\frac{\lambdabar_{\rm C}}{w_0}\Bigr)^4 \, \frac{F_0}{\cal P}\,. \label{eq:NperpbyNF}
\end{align}
Here, we have divided both sides of \Eqref{eq:NperpbyN} by a factor of ${\cal P}$. As the combination $N{\cal P}$ corresponds to the number of background photons against which the signal $N_\perp$ has to be distinguished in experiment, \Eqref{eq:NperpbyNF} counts the maximum attainable number of polarization-flipped signal photons per background photon.
Hence, the criterion to obtain one signal photon per one background photon is 
\begin{align}
\frac{4\alpha^4}{25(3\pi)^\frac{3}{2}}\,\Bigl(\frac{W}{m_e}\frac{\omega}{m_e}\Bigr)^2\Bigl(\frac{\lambdabar_{\rm C}}{w_0}\Bigr)^4 \, \frac{F_0}{\cal P}=1\,. \label{eq:discern}
\end{align}
At HiBEF, the maximal achievable value for \Eqref{eq:NperpbyNF} is $N_\perp/(N{\cal P})\simeq1/40$; cf. also Fig.~\ref{fig:full}.

For completeness, we note that for $\beta\ll1$ the signal photons and the probe photons traversing the strong-field region without interaction feature similar far-field divergences \cite{Karbstein:2018omb}. Therefore, the ratio of signal to background photons is actually well-described by \Eqref{eq:NperpbyNF} not only for the integrated photon numbers but in fact also for the differential photon numbers associated with arbitrary emission directions.

\subsection{Discernible polarization-flipped signal photons}\label{subsec:b}

In a next step, we study the discernible number of polarization-flipped signal photons, i.e., the number of polarization-flipped signal photons scattered outside the forward cone of the probe beam.
Signal photons are {\it discernible} from the background if they fulfill the criterion
\begin{align}
\frac{{\rm d}N_{\perp}}{\vartheta\,{\rm d}\vartheta}\geq {\cal P}\,\frac{{\rm d}N}{\vartheta\,{\rm d}\vartheta}\, \label{eq:discerncrit}
\end{align}
for the differential numbers of polarization-flipped signal photons~\eqref{eq:d2N} and the photons available for probing~\eqref{eq:decay}.

As the far-field divergence of the signal is generically larger or equal the far-field divergence of the probe beam \cite{Karbstein:2018omb}, for ${\rm d}N_\perp/(\vartheta\,{\rm d}\vartheta)|_{\vartheta=0}<{\cal P}\,{\rm d}N/(\vartheta\,{\rm d}\vartheta)|_{\vartheta=0}$,
there is always an angle $\vartheta_=$ such that signal photons emitted under an angle of $\vartheta\geq\vartheta_=$ fulfill the criterion~\eqref{eq:discerncrit}.
This angle can be determined analytically and reads
\begin{align}
\vartheta_{=}^{2}&=\frac{-2}{(\omega\beta w_0)^2\bigl(1-\frac{1}{1+2\beta^2}\sqrt{\frac{F_\beta}{F_0}}\bigr)}\label{eq:angleperp}\\\nonumber
&\quad\times\ln{\biggl(\frac{4\alpha^4}{25(3\pi)^{\frac{3}{2}}}\Bigl(\frac{W}{m_e}\frac{\omega}{m_e}\Bigr)^2\Bigl(\frac{\lambdabar_{\rm C}}{w_0}\Bigr)^4
 \frac{F_\beta/{\cal P}}{(1+2\beta^2)^2}\biggr)}\,.   
\end{align}

Integrating \Eqref{eq:d2N} over all angles $\vartheta\geq\vartheta_=$, we arrive at the following expression for the discernible number of signal photons  
\begin{align}
 N_{\perp >}
 &\simeq
 {\cal P} N\,
(1+2\beta^2) \sqrt{\frac{F_0}{F_\beta}}\label{eq:Ndiscern}\\\nonumber &\quad\times \biggl(\frac{4\alpha^4}{25(3\pi)^{\frac{3}{2}}}\Bigl(\frac{W}{m_e}\frac{\omega}{m_e}\Bigr)^2\Bigl(\frac{\lambdabar_{\rm C}}{w_0}\Bigr)^4
 \frac{F_\beta/{\cal P}}{(1+2\beta^2)^2}\biggr)^{\kappa} \,,
\end{align}
with exponent
\begin{equation}
\kappa=\frac{1}{1-\frac{1}{1+2\beta^2}\sqrt{\frac{F_\beta}{F_0}}}\label{eq:kappa}\,.
\end{equation}
As these signal photons are discernible from the background by definition, we are generically interested directly in their number and not in their number normalized by the number of background photons.
For later reference, we nevertheless also introduce the number of background photons scattered into the angular regime $\vartheta\geq\vartheta_=$ as $N_>$. This number follows readily from \Eqref{eq:decay}.

\begin{figure}
\center
\includegraphics[width=0.45\textwidth]{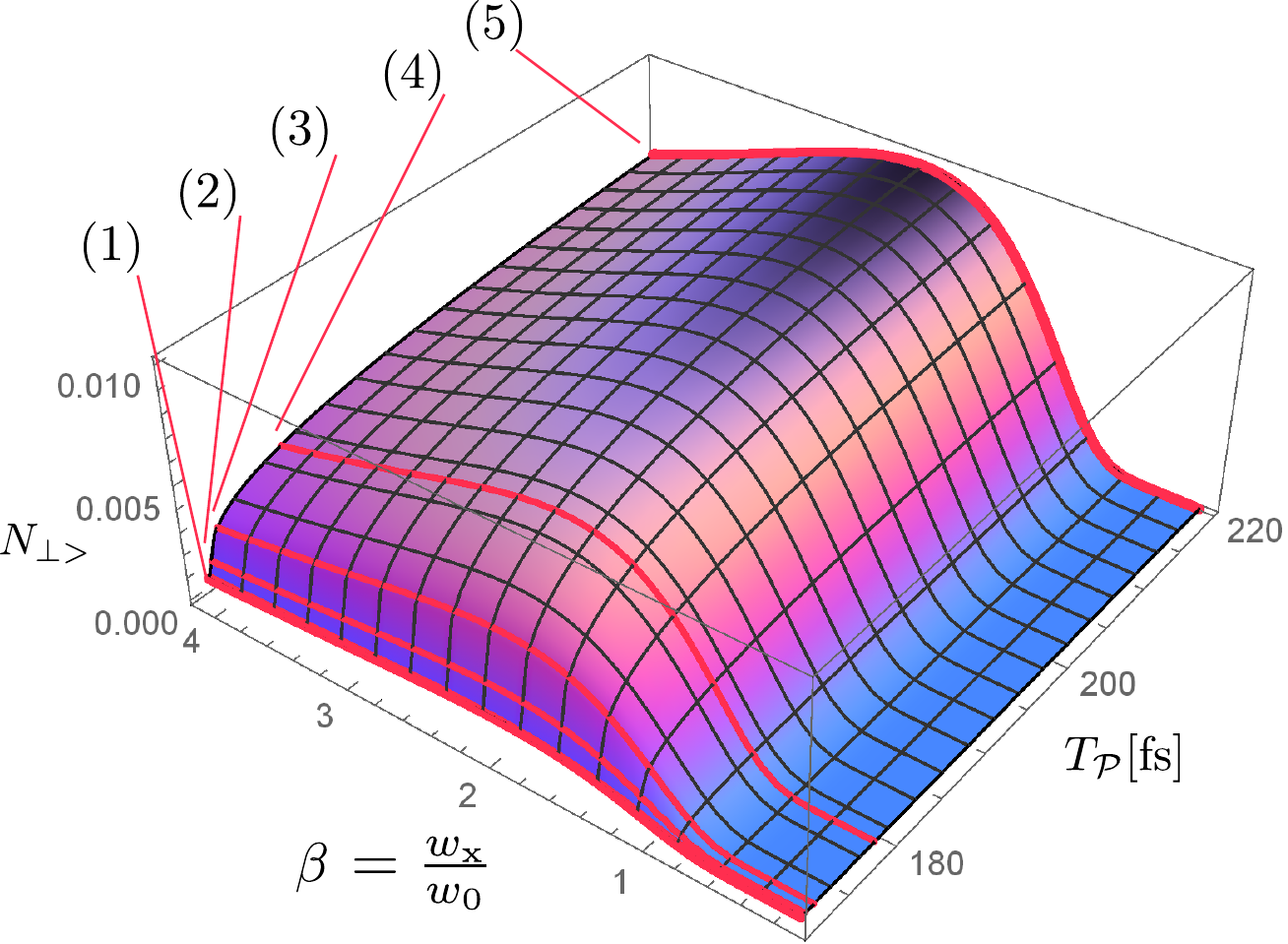}
\caption{
Dependence of the discernible signal $N_{\perp>}$ on the probe pulse duration $T_{\cal P}$ and probe waist $\beta=w_{\rm x}/w_0$ for $\omega=12914\,{\rm eV}$ and ${\cal P}=1.4 \times 10^{-11}$. 
The pump laser parameters are those available at HiBEF: $W=10\,{\rm J}$, $\tau=42\,{\rm fs}$, $\lambda=800\,{\rm nm}$, $w_0=1.7\upmu{\rm m}$.
The maximum number of $N_{\perp>}\simeq0.01/$shot ($N_{\perp >}\simeq36/$hour for a repetition rate of $1\,{\rm Hz}$) is reached for (5): $T^\text{FWHM}=107\, {\rm fs}$ and $w_{\rm x}\simeq 2.1 \,w_0$. The red contours (1)-(5) mark the results obtained for the probe pulse durations listed in Tab.~\ref{tab:times}; for intermediate times we use a smooth monotonic interpolation $N(T)$ of the these values.}
\label{fig:dis}
\end{figure}

The parameter dependence of the discernible signal is more complicated than that of the integrated number discussed in Sec.~\ref{subsec:a}.
In Fig.~\ref{fig:dis} we highlight the dependence of the discernible signal on both the pulse duration $T_{\cal P}$ and the waist $w_{\rm x}$ of the x-ray probe.
The growth of $N_{\perp>}$ with $T_{\cal P}$ can be explained by the increase of the number of XFEL photons available for probing, i.e., is due to the non-trivial correlation between $N$ and $T$; see Tab.~\ref{tab:times}.
This implies that in contrast to the behavior observed for the ratio $N_\perp/({\cal P}N)$ in Sec.~\ref{subsec:a}, for the discernible signal the optimal choice for the pulse duration is the largest one.
Recall, that the ratio $N_\perp/(N{\cal P}$) is the decisive quantity for experiments aiming at measuring integrated signal photon numbers without resorting to an explicit discernibility criterion.
Apart from that, the maximal number of discernible signal photons is achieved for a finite probe waist, and not for $\beta\to0$ as for $N_\perp/({\cal P}N)$.

The high quality of our approximation, even allows us to find the optimal x-ray waist $w_{\rm x}^{\rm opt}=\beta_{\rm opt}w_0$ by differentiation for $\beta$. This results in the condition
\begin{align}
 \frac{4\alpha^4}{25(3\pi)^{\frac{3}{2}}}\Bigl(\frac{W}{m_e}\frac{\omega}{m_e}\Bigr)^2\Bigl(\frac{\lambdabar_{\rm C}}{w_0}\Bigr)^4
 \frac{F_0}{{\cal P}}
 =\frac{1}{\chi^2}\exp{\Bigl(\chi-\frac{1}{\chi}\Bigr)}
  \,, \label{eq:OpWaist}
\end{align}
where we introduced
\begin{align}
    \chi=\frac{1}{1+2\beta^2}\sqrt{\frac{F_\beta}{F_0}}\,\bigg|_{\beta=\beta_{\rm opt}}\in \,(0,1]\,.\label{eq:chi}
\end{align}
Note, that $\chi$ decreases with growing $\beta_{\rm opt}$.     

For $\chi\in(0,1]$ also the right-hand side of \Eqref{eq:OpWaist} takes values from the interval $(0,1]$. Hence, the necessary criterion for the existence of an optimal x-ray waist maximizing the effect is 
\begin{align}
 \frac{4\alpha^4}{25(3\pi)^{\frac{3}{2}}}\Bigl(\frac{W}{m_e}\frac{\omega}{m_e}\Bigr)^2\Bigl(\frac{\lambdabar_{\rm C}}{w_0}\Bigr)^4
 \frac{F_0}{{\cal P}} \label{eq:parameter}
 \leq1\,.
\end{align}
Equation~\eqref{eq:OpWaist} has no solution in the complementary parameter regime.
From the fact that the right-hand side of \Eqref{eq:OpWaist} scales inversely with $\beta_{\rm opt}$, we can moreover infer that the larger \Eqref{eq:parameter}, the smaller the optimal x-ray waist.
When the equality holds, i.e., the criterion~\eqref{eq:discern} is met, $\beta_{\rm opt}\to0$. In this case all the photons become discernible.  

Figure~\ref{fig:Wdep} exemplifies the dependence of the discernible signal $N_{\perp >}$ on the probe waist for different pump pulse energies $W$ and polarization purities $\cal P$:
\begin{figure}
\center
\includegraphics[width=0.5\textwidth]{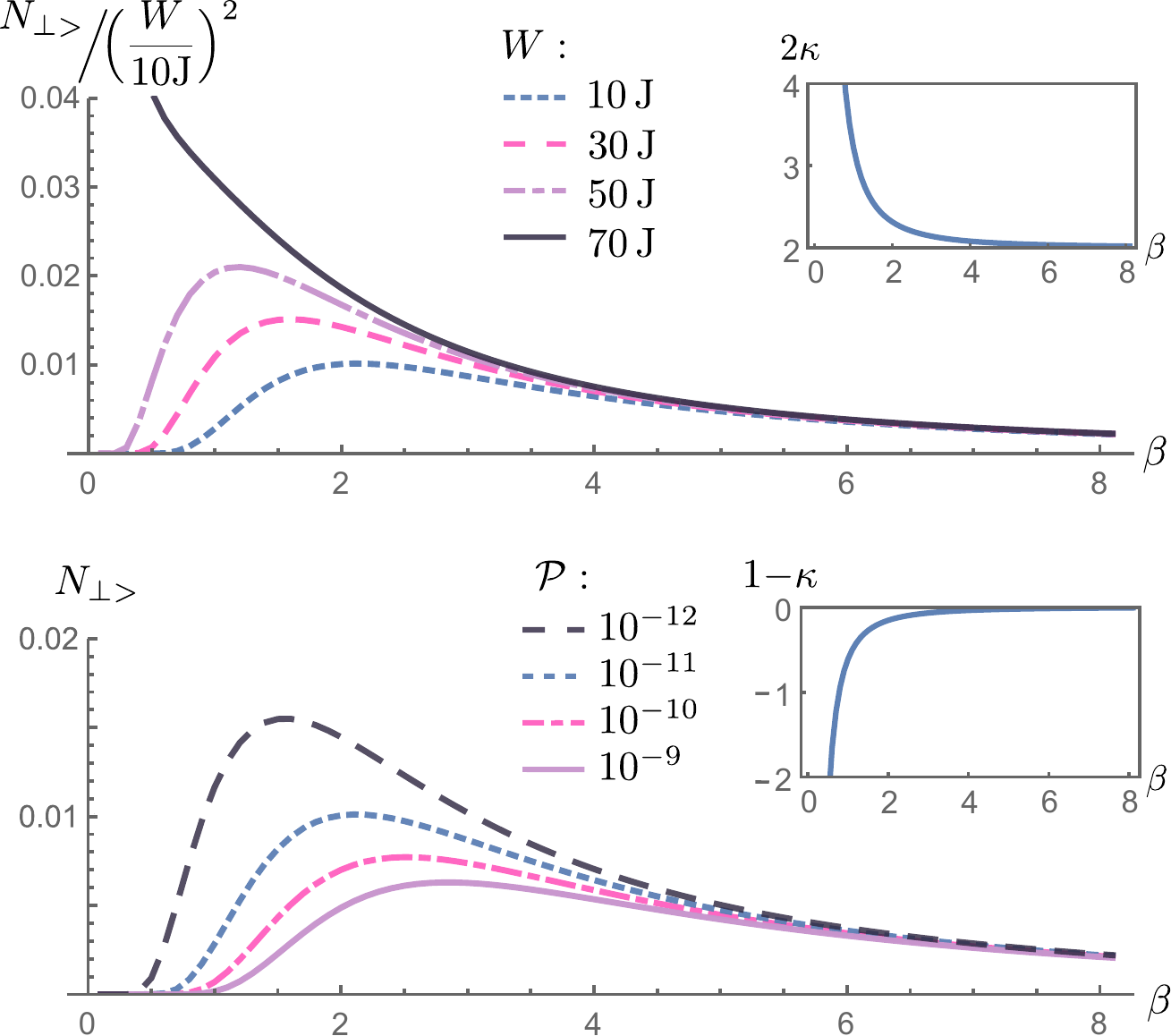}
\caption{Upper panel: dependence of $N_{\perp >}$ (normalized) on the pump pulse energy $W$ for a polarization purity of ${\cal P}=1.4\times10^{-11}$. The inset highlights the $\beta$-dependence of the exponent $2\kappa$ of the pump pulse energy; recall that $N_{\perp>}\sim W^{2\kappa}$.
Lower panel: dependence of $N_{\perp >}$ on the polarization purity ${\cal P}$ for $W=10\,{\rm J}$.
The inset highlights the $\beta$-dependence of the exponent $1-\kappa$ of the polarization purity; recall that $N_{\perp>}\sim {\cal P}^{1-\kappa}$.
The XFEL pulse duration is fixed to option (5) in Tab.~\ref{tab:times}. The other parameters are fixed to those available at HiBEF: $\omega=12914\,{\rm eV}$, $\tau=42\,{\rm fs}$, $\lambda=800\,{\rm nm}$, $w_0=1.7\upmu{\rm m}$.
To obtain a feeling for the vertical scales in the upper and lower panels, note that the blue dashed lines in both panels are for the same other.}
\label{fig:Wdep}
\end{figure}
for pump pulse energies meeting the criterion~\eqref{eq:parameter}, $N_{\perp >}$ features a maximum at a finite value of $\beta$. The location $\beta_{\rm opt}$ of the maximum allows us to infer the optimal waist $w_{\rm x}^{\rm opt}$. From Fig.~\ref{fig:Wdep} it is clear that the optimal waist decreases with increasing pump energy.

Moreover, specifically the inset in the upper panel of Fig.~\ref{fig:Wdep} illustrates that the pump energy dependence of the effect is enhanced for decreasing $\beta$. 
Recall, that Eqs.~\eqref{eq:Ndiscern} and \eqref{eq:kappa} imply that $N_{\perp>}\sim W^{2\kappa}$.
For the HiBEF parameters given above ($W=10\,{\rm J}$), the optimal probe waist is $w^{\rm opt}_{\rm x}\simeq 2.1 w_0$. The exponent governing the pump energy dependence associated with this value is $2\kappa\simeq2.3$.
Hence, for the choice of $w_{\rm x}=w^{\rm opt}_{\rm x}\simeq 2.1 w_0$ this exponent governs the behavior of the signal $N_{\perp>}$ with regard to moderate changes of the pump pulse energy at HiBEF.

On the other hand, the inset in the lower panel of Fig.~\ref{fig:Wdep} highlights that the dependence of $N_{\perp>}\sim {\cal P}^{1-\kappa}$ on the polarization purity is also enhanced for decreasing $\beta$.
For the HiBEF parameters with $w_{\rm x}=w^{\rm opt}_{\rm x}\simeq 2.1 w_0$, the value of the exponent governing the dependence of the signal photon number on $\cal P$ is $1-\kappa\simeq-0.14$.

We note that in the parameter regime relevant for an vacuum birefringence experiment at XFEL we generically have $1-\kappa\ll 2\kappa$, such that the dependence of $N_{\perp>}$ on $\cal P$ is much weaker than that on $W$. This immediately suggests that an increase of the pump pulse energy by a given factor is more effective for enhancing the signal than an improvement of the polarization purity by the same factor.

Finally, in an attempt to compare the observable $N_{\perp>}$ analyzed in the present section with $N_\perp/(N{\cal P})$ discussed in Sec.~\ref{subsec:a}, we consider the analogous ratio for the number of signal photons per background photon scattered outside the discernibility angle $\vartheta_{=}$,
\begin{align}
\frac{N_{\perp>}}{N_> {\cal P}}=(1+2\beta^2)\sqrt{\frac{F_0}{F_\beta}}\,. \label{eq:NperpbyNFdis}
\end{align}
This ratio counts the number of discernible polarization-flipped signal photons per background photon. It depends only on geometric properties of the experimental setup.
For the parameters available at HiBEF and a probe waist of $w_{\rm x}=w^{\rm opt}_{\rm x}\simeq 2.1 w_0$ we have $N_{\perp>}/(N_>{\cal P})\simeq8$.
A comparison with the value of $N_\perp/(N{\cal P})\simeq1/40$ obtained below \Eqref{eq:discern} exemplifies the huge enhancement potential of the study of discernible signals with respect to the integrated ones. 

\subsection{Total number of discernible signal photons}\label{subsec:c}

Finally, we study the discernible part of the total number of signal photons attainable in a polarization insensitive measurement. By definition these photons fulfill
\begin{align}
\frac{{\rm d}N_{\rm tot}}{\vartheta\,{\rm d}\vartheta}\geq \frac{{\rm d}N}{\vartheta\,{\rm d}\vartheta}\,.
\end{align}
Due to the fact that generically ${\rm d}N_{\rm tot}/(\vartheta\,{\rm d}\vartheta)|_{\vartheta=0}<{\rm d}N/(\vartheta\,{\rm d}\vartheta)|_{\vartheta=0}$, there is always an angle $\vartheta_=$ from which onward this criterion is met.
The explicit result for $\vartheta_=$ can be extracted from \Eqref{eq:decay} and the first line of \Eqref{eq:d2N}. It reads
\begin{align}
\vartheta_{=}^{2}&=\frac{-2}{(\omega\beta w_0)^2\bigl(1-\frac{1}{1+2\beta^2}\sqrt{\frac{F_\beta}{F_0}}\bigr)}\label{eq:angleperptot}\\\nonumber
&\times\ln{\biggl(\frac{784\alpha^4}{225(3\pi)^{\frac{3}{2}}}\Bigl(\frac{W}{m_e}\frac{\omega}{m_e}\Bigr)^2\Bigl(\frac{\lambdabar_{\rm C}}{w_0}\Bigr)^4
 \frac{F_\beta}{(1+2\beta^2)^2}\biggr)}\,.   
\end{align}
Upon integration of ${\rm d}N_{\rm tot}/(\vartheta\,{\rm d}\vartheta)$ over all polar angles fulfilling $\vartheta\geq \vartheta_{=}$, we arrive at the following expression for the discernible signal,
\begin{align}\label{eq:Ntotdiscern}
 N_{\rm tot >}
 &\simeq
  N\,
(1+2\beta^2) \sqrt{\frac{F_0}{F_\beta}}\\\nonumber&\times \biggl(\frac{784\alpha^4}{225(3\pi)^{\frac{3}{2}}}\Bigl(\frac{W}{m_e}\frac{\omega}{m_e}\Bigr)^2\Bigl(\frac{\lambdabar_{\rm C}}{w_0}\Bigr)^4
 \frac{F_\beta}{(1+2\beta^2)^2}\biggr)^\kappa \,,
\end{align}
with exponent $\kappa$ defined in \Eqref{eq:kappa}.
Obviously, the parameter dependence of $N_{{\rm tot}>}$ is very similar to the one inferred for $N_{\perp>}$ in Sec.~\ref{subsec:b}.
In fact, \Eqref{eq:Ntotdiscern} follows from \Eqref{eq:Ndiscern} upon substituting ${\cal P}\to1$ and $\alpha^4\to196\alpha^4/9$.

\begin{figure}
\center
\includegraphics[width=0.45\textwidth]{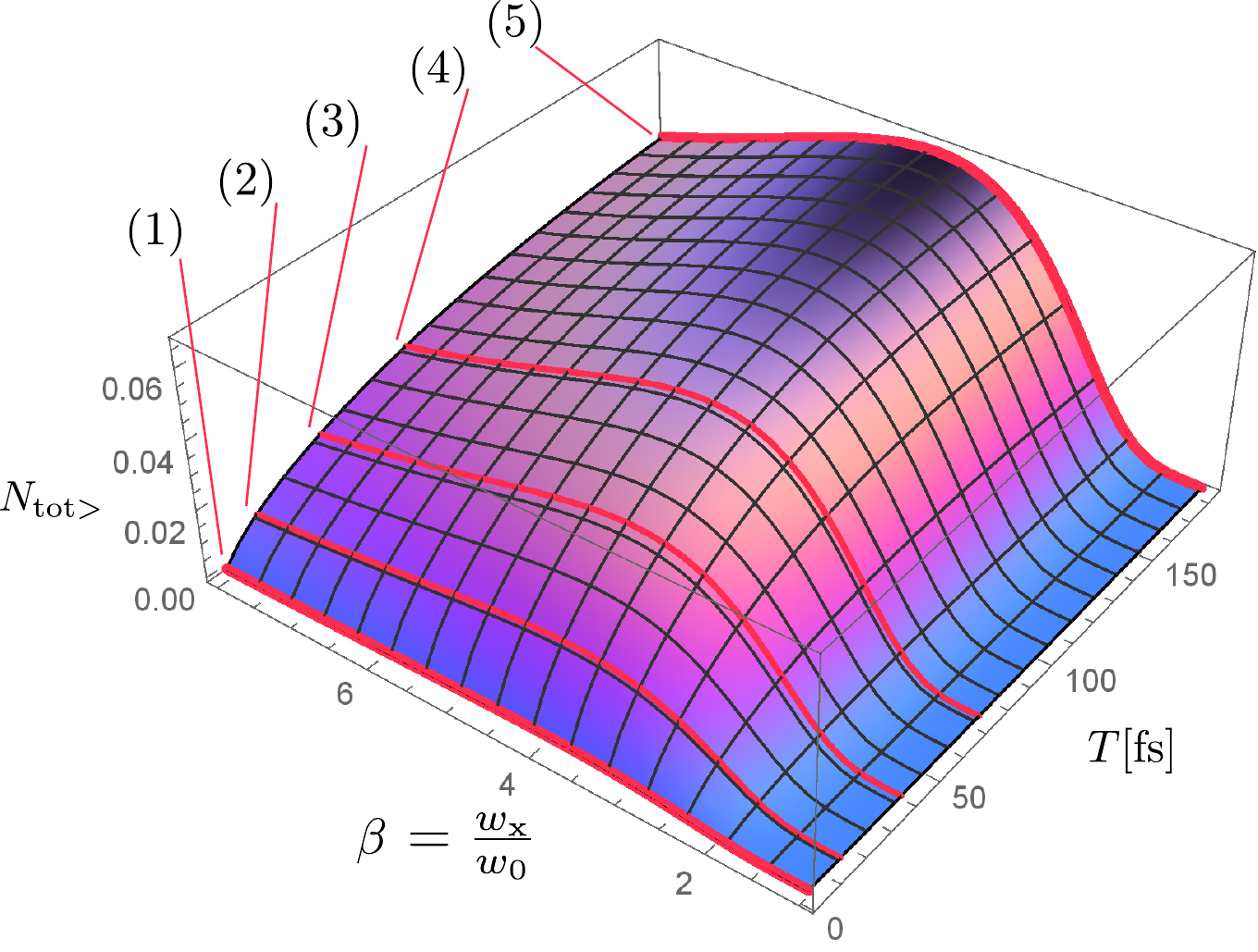}
\caption{Dependence of the number of discernible signal photons $N_{{\rm tot}>}$ on the probe pulse duration $T$ and the focal spot ratio $\beta=w_{\rm x}/w_0$ for a photon energy of $\omega=12914\,{\rm eV}$.
The pump laser parameters are those available at HiBEF: $W=10\,{\rm J}$, $\tau=42\,{\rm fs}$, $\lambda=800\,{\rm nm}$, $w_0=1.7\upmu{\rm m}$.
The maximum number of $N_{{\rm tot}>}\simeq0.07/$shot ($N_{{\rm tot}>}=241/$hour for a repetition rate of $1\,{\rm Hz}$) is reached for (5): $T^\text{FWHM}=107\, {\rm fs}$ and $w_{\rm x}\simeq 4.5 \,w_0$.
The red contours (1)-(5) mark the results obtained for the probe pulse durations listed in Tab.~\ref{tab:times}; for intermediate times we use a smooth monotonic interpolation $N(T)$ of the these values.}
\label{fig:tot}
\end{figure}
As no high-definition polarimetry is required for the detection of this quantity, the probe pulse durations $T$ available for measuring $N_{{\rm tot}>}$ are those given in the second column of Table \ref{tab:times}.

Figure~\ref{fig:tot} highlights the dependence of $N_{{\rm tot}>}$ on the probe pulse duration $T$ and probe waist $w_{\rm x}$ for the parameters available at HiBEF. 
A comparison with Fig.~\ref{fig:dis} unveils that the maximum value for $N_{{\rm tot}>}$ is about a factor of $7$ larger than the maximum value for $N_{\perp>}$.
Also note that the substantial difference in the values of $w_{\rm x}$ for these maxima:
for the polarization-flipped signal we have $w_{\rm x}\simeq2.1w_0$, and for the discernible signal studied here $w_{\rm x}\simeq4.5w_0$.

For the present observable, the analogue of the condition~\eqref{eq:OpWaist} for the optimal waist is given by
\begin{align}
 \frac{784\alpha^4}{225(3\pi)^{\frac{3}{2}}}\Bigl(\frac{W}{m_e}\frac{\omega}{m_e}\Bigr)^2\Bigl(\frac{\lambdabar_{\rm C}}{w_0}\Bigr)^4
 F_0
 =\frac{1}{\chi^2}\exp{\bigl(\chi-\frac{1}{\chi}\bigr)}
  \,, \label{eq:OpWaistTot}
\end{align}
with $\chi$ defined in \Eqref{eq:chi}.
For the HiBEF parameters this condition predicts an optimal probe waist of $w_{\rm x}^{\rm opt}\simeq4.5 w_0$ in accordance with the value inferred from Fig.~\ref{fig:tot}.
Noteworthy, for $N_{{\rm tot}>}$ the optimal waist is very stable with respect to variations of all experimental parameters.

At the same time, the ratio counting the numbers of signal photons per background photon reads
\begin{align}
\frac{N_{\rm tot >}}{N_>}=(1+2\beta^2)\sqrt{\frac{F_0}{F_\beta}} \,. \label{eq:NtotbyNFdis}
\end{align}
Interestingly, this result exactly matches \Eqref{eq:NperpbyNFdis}.
For the HiBEF parameters and $w_{\rm x}=w_{\rm x}^{\rm opt}\simeq4.5w_0$ we find $N_{{\rm tot}>}/N_>\simeq32.6$.
\begin{figure}
\center
\includegraphics[width=0.48\textwidth]{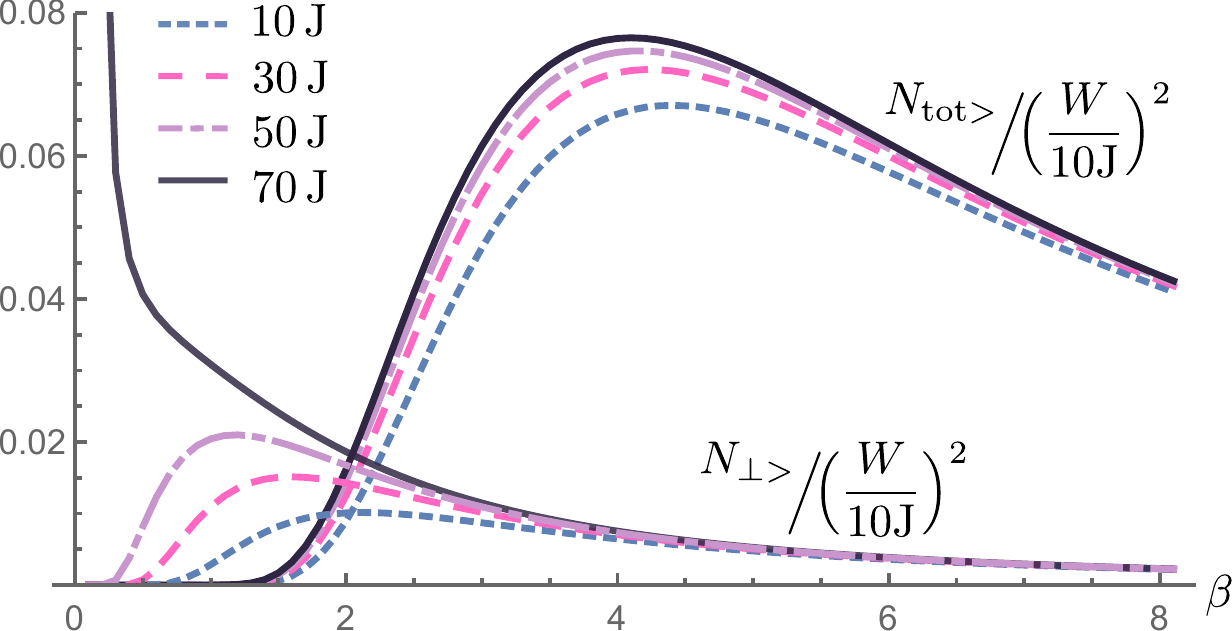}
\caption{Dependence of $N_{\perp >}$ and $N_{\rm tot >}$ (normalized) on the pump pulse energy $W$. The XFEL pulse duration is fixed to option (5) in Tab.~\ref{tab:times}.
The other parameters are fixed to those available at HiBEF: $\omega=12914\,{\rm eV}$, ${\cal P}=1.4\times10^{-11}$, $\tau=42\,{\rm fs}$, $\lambda=800\,{\rm nm}$, $w_0=1.7\upmu{\rm m}$. The curves for $N_{\perp >}$ agree with those plotted in the upper panel of Fig.~\ref{fig:Wdep}.}
\label{fig:TotDisW}
\end{figure}

The dependence $N_{{\rm tot}>}\sim W^{2\kappa}$ matches the one for $N_{\perp>}$.
However, as noted above, in the present case the value of the optimal probe waist is larger. Hence, for the HiBEF parameters we arrive at a slightly different value of the exponent as in Sec.~\ref{subsec:b}: for the present observable the exponent is given by $2\kappa\simeq2.06$.
See Fig.~\ref{fig:TotDisW} for a more detailed study of the pump energy dependence of both observables.

\begin{figure}
\center
\includegraphics[width=0.48\textwidth]{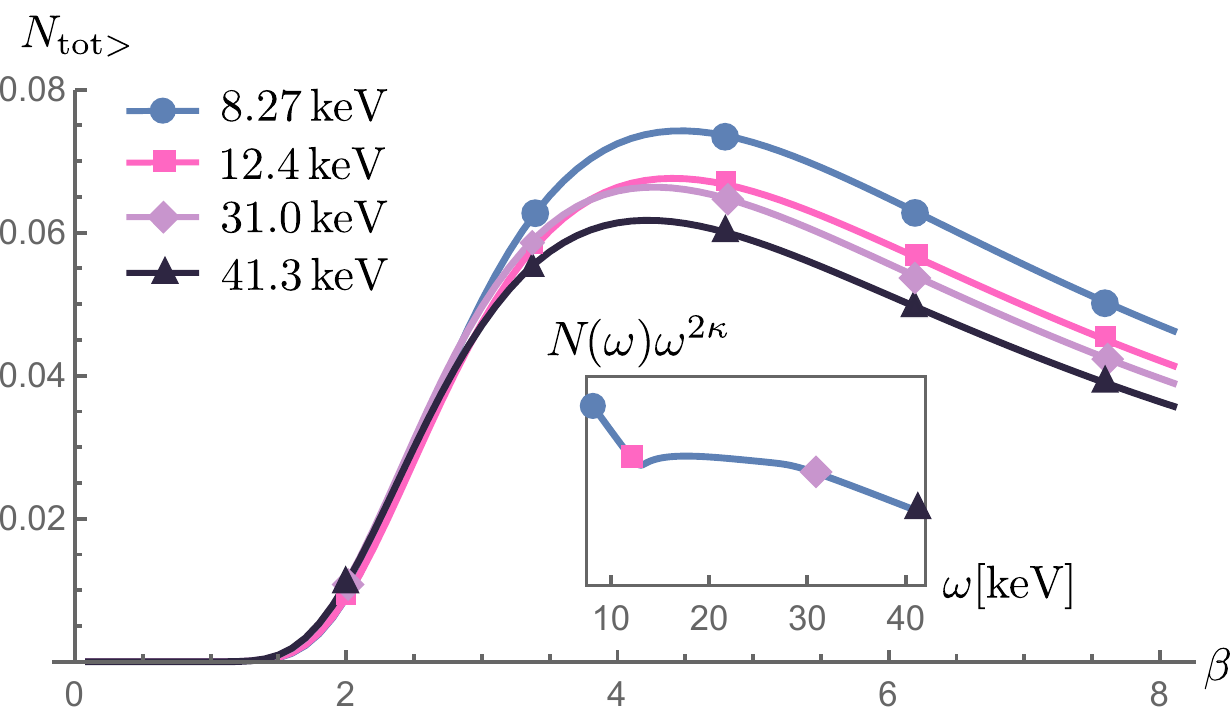}
\caption{Dependence of the discernible signal $N_{{\rm tot}>}$ on $\beta=w_{\rm x}/w_0$ for $T^{\rm FWHM}=107\,{\rm fs}$ and different probe photon energies $\omega$, but fixed other parameters of HiBEF: $W=10\,{\rm J}$, $\tau=42\,{\rm fs}$, $\lambda=800\,{\rm nm}$, $w_0=1.7\upmu{\rm m}$.
The inset shows the scaling of $N_{\rm tot >}$ at the optimal waist of $\beta_{\rm opt}\simeq4.5$ with $\omega$ for $2\kappa\simeq2.06$; recall that $N_{\rm tot >}\sim N(\omega)\omega^{2\kappa}$.}
\label{fig:TotDisOmega}
\end{figure}

Because no high-definition polarimetry is needed for the measurement of $N_{{\rm tot}>}$, we are essentially free in choosing the probe photon energy $\omega$.
On the other hand, the number of photons available for probing at XFEL depends on the chosen value of $\omega$, such that $N\to N(\omega)$; see Tabs.~C.1-C.6 of Ref.~\cite{Schneidmiller:95609} for the explicit values for various values of $\omega$.
From \Eqref{eq:Ntotdiscern} one can see that $N_{{\rm tot} >}$ scales with $\omega$ as $N_{\rm tot >}\sim N(\omega)\omega^{2\kappa}$, where $2\kappa\simeq2.06$; see the inset in Fig.~\ref{fig:TotDisOmega} highlighting this dependence for the European XFEL.

Figure~\ref{fig:TotDisOmega} demonstrates that $N_{{\rm tot}>}$ is essentially insensitive to changes of the probe photon energy in the wide range of $8.27\,{\rm keV}\leq\omega\leq41.3\,{\rm keV}$. At the European XFEL, the maximum value of the discernible signal would be obtained for the smallest considered frequency of $\omega=8.27\,{\rm keV}$. However, the gain relative to the photon energy of $\omega=12914\,{\rm eV}$ assumed throughout this work is insignificant.

\section{Conclusions and Outlook}
\label{sec:Concls}

In the present work, we have constructed an improved analytic approximation for the differential numbers of signal photons encoding the experimental signature of vacuum birefringence and diffraction in the head-on collision of an XFEL probe with a high-intensity laser pump.
Our approximation allows for quantitatively accurate studies of the prospective signals; the relative deviation to the corresponding exact, numerical results is on the $1\%$ level.
One of the key advantages of our analytical results is the possibility of a direct study of the behavior of the signal under the variation of various experimental parameters.

Focusing on the experimental parameters available at HiBEF at the European XFEL, and explicitly taking into account the non-trivial dependence of the number of XFEL photons available for probing on the XFEL pulse duration and photon energy, we determined the optimal choices for the parameters in experiment such as to maximize the signal.
To this end, we analyzed three distinct experimental observables, namely the integrated number of signal photons $N_\perp$, the discernible number of polarization-flipped signal photons $N_{\perp>}$ and the discernible number of signal photons attainable in a polarization insensitive measurement $N_{{\rm tot}>}$.
We showed that, due to the distinct parameter dependencies, the optimization of each of these observables requires different choices of the beam waist of the probe.
Moreover, we demonstrated that maximizing the integrated (discernible) signal photon number requires choosing the minimal (maximal) probe pulse duration.

Even though throughout the present study we explicitly limited ourselves to optimal laser pulse collisions at zero impact parameter, we expect only minor corrections when accounting for finite impact parameters $r_0\lesssim w_0$, which amount to typical beam jitters of tightly focused optical high-intensity lasers. In Sec.~\ref{sec:results}, we inferred that the optimal discernible signals are obtained for $\beta_{\rm opt}>1\leftrightarrow w_{\rm x}>w$, with $\beta_{\rm opt}=2.1$ ($4.5$) for the $\perp$-polarized (total) signal. In this parameter regime \Eqref{eq:impact} predicts the reduction of the signal to be approximately governed by $\approx{\rm e}^{-(r_0/w_{\rm x})^2}$. Hence, for impact parameters $r_0\leq w_0$ the signal is at most reduced by a factor of $\approx{\rm e}^{-1/\beta^2}$ and the optimal parameters should essentially not be modified. On the other hand, the optimal result for the integrated number of signal photons is achieved for $\beta\ll1$.
In this limit, \Eqref{eq:impact} predicts a reduction of the signal by at most a factor of $\approx{\rm e}^{-4}$ for $r_0\leq w_0$ independent of  $w_{\rm x}$.
In turn, also this finding should not be modified for collisions under a finite impact parameter.

We expect our results to be of large relevance for the identification of the optimal parameters for experiments aiming at the detection of vacuum birefringence in XFEL/high-intensity laser setups, particularly the one put forward at HiBEF.

\acknowledgments

The work has  been  funded  by  the  Deutsche Forschungsgemeinschaft  (DFG)  under  Grant  No. 416607684 within the Research Unit FOR2783/1 and was supported by the state assignment of the Ministry of Science and Higher Education of the Russian Federation (theme No. AAAA-A20-120090990006-0).  Moreover, E.A.M. would like to thank the Helmholtz Institute Jena for hospitality and support. Both authors are grateful to K.~S.~Schulze and B.~Marx-Glowna for helpful discussions.

%\appendix
%\section{The First Appendix}

\end{document}